# Revealing mesoscale bubble and particle dynamics in ultrasound-driven multiphase fluids by ultrafast synchrotron X-ray radiography and hybrid modelling

Ling Qin[1,2], Kang Xiang[1], Iakovos Tzanakis[3], Dmitry Eskin[4], Samuel Clark[5], Kamel Fezzaa[5], Jiawei Mi[1,*]

1. School of Engineering and Technology, University of Hull, East Yorkshire, HU6 7RX, UK
2. Centre of Innovation for Flow Through Porous Media, Department of Energy and Petroleum Engineering, University of Wyoming, Laramie, WY 82071, USA
3. Department of Mechanical Engineering and Mathematical Sciences, Oxford Brookes University, Oxford, OX3 0BP, UK
4. Brunel Centre for Advanced Solidification Technology, Brunel University London, Uxbridge, UB8 3PH, UK
5. The Advanced Photon Source, Argonne National Laboratory, Argonne, 60439, USA

* Corresponding author: Jiawei. Mi   j.mi@hull.ac.uk

## Abstract:

Multiphase fluid flows comprising of mesoscale solid particles, liquid droplets, and/or gas bubbles are common in both natural and man-made systems, but quantifying the energy transfer is challenging due to complex bubble-particle interactions. In this study, we used ultrafast synchrotron X-ray imaging to study the mesoscale dynamic interactions among ultrasonic cavitation bubbles and hydrophobic particles/clusters. Critical dynamic information and data were extracted from the vast amount of X-ray images and then fed into the hybrid analytical-numerical model for calculating the energy transfer from the oscillating bubble and the imploding bubble to the nearby hydrophobic particles. Using the Ni spherical microparticles as an example, at bubble oscillation ~16% (80–320 nJ) of the local energy was transferred to the particle. At bubble implosion, the transferred energy increased ~26% (0.135–1.09 µJ). Local energy transfer occurred on timescales of ~1 µs to 1 ms and length scales of 1 µm to 1 mm. Within each ultrasound cycle, kinetic and potential energy underwent complex exchanges, with local energy exhibiting a stepwise decay at the end of each cycle. The transferred energy was mainly consumed for enabling highly efficient particle dispersion. This research provides quantitative insights into optimizing hydrophobic nanomaterial dispersion and has broader implications for interfacial energy transfer processes such as making suspensions, composite materials and exfoliated 2D materials.

**Main**

Fluid flows consisting of mesoscale multiple discrete phases of different sizes (i.e., solid particles, liquid droplets and/or gas bubbles) are ubiquitous in naturally-occurred and man-made fluid systems[1-3], e.g., in air, water or other types of fluids containing multiple phases[4]. Dynamic interactions among the discrete phases and those between the discrete phases and the continuous fluid are governed by complex, nonlinear mass and energy transfer processes occurring across multi-scale spatiotemporal domains. Over the past four decades, significant progresses have been made in measuring and understanding the hydrodynamic behaviour of multiphase flow due to the advances of sophisticated experimental and computational techniques, for example, laser Doppler anemometry[5], particle image velocimetry[6], computer-automated radioactive particle tracking[7,8], and optical bubble probes[9]. However, until now, at the mesoscale (i.e., length scales ranging from tens of nm to hundreds of µm, and time scales from ns to µs), it is still a great experimental challenge to observe and measure in real-time or operando conditions the dynamic phenomena occurring at the above spatiotemporal scale[10,11]. The even more challenging and least understood scientific cases are multiphase dynamic interactions under the influence of ultrasonic waves[12,13]. The chaotic ultrasonic bubble implosion may produce impulsive energy in the range of 10-100 millijoules (depending on the bubble size and local acoustic pressure[14]), acting on the multiphases that may have distinct differences or disparity in size, morphology, bulk (or surface) properties[15-18].

To study the dynamic interactions among microbubbles and particles, high speed optical imaging techniques have been widely used. In 1933, Wark[19] and Kabanow[20] made the 1st attempt, studied a static bubble's interaction with a particle and calculated the force balance at the bubble interface. However, the model they used was not suitable for dynamic bubbles. For dynamic bubbles, powerful laser was often used to create a controllable sequence of bubble near some microparticles in a flow medium for facilitating experimental investigations. For example, Poulain et al.[21] studied a single hydrophilic glass particle interacting with an oscillating bubble and found that the particle's velocity gained after the interaction is related to its distance away from the bubble (an inverse fourth-power law). Wu *et al.*[22] also observed the bouncing off and ejecting behaviour of hydrophilic particles by expanding bubble boundary. Classic force balance models were often used to calculate the particle's trajectory caused by bubble oscillation or implosion as well as the energy transfer during the interaction. In most cases, conventional optical imaging is not an ideal technique for imaging the interaction between hydrophobic particles (or clusters) and bubbles, as these particles often "mangle or tangle" together with the bubbles' interfaces, creating multi-directional light scattering effects which "smear or blur" the bubble interfaces in the captured images[23]. In this aspect, the third-generation synchrotron X-ray facilities started in the mid-1990s and the X-ray Free Electron Lasers (XFELs) have provided the high-brilliance X-rays sources which enabled unprecedented capabilities and advantages over optical imaging (because of virtually no scattering effect across the particle-bubble boundaries) for studying in-situ and in operando conditions the complex multiphase flow dynamics[24-26]. Unfortunately, until now, systematic measurements made in operando conditions and high-fidelity simulations of these multiphase interactions, especially accurate quantification of the associated mass and energy transfers have not been realised for many scientifically and technologically important flow systems, particularly lack of systematic studies when ultrasonic waves are present in the multiphase flows.

Here, we used ultrafast synchrotron X-ray phase-contrast imaging to observe the oscillation and movement of ultrasonic bubbles, as well as their dynamic interactions with hydrophobic microparticles (Ni spheres and $MoS_2$ clusters in this study) under operando conditions. We also developed a novel hybrid analytical-numerical modelling methodology based on the key information extracted from the X-ray images to calculate the kinetic energy and potential energy transfer between the particles and the bubbles in oscillation or at the instant of bubble implosion. The new scientific understandings and high-fidelity modelling of these mesoscale dynamics under ultrasound are valuable for design optimisation and process control in a vast amount of multiphase fluid systems with the presence of ultrasounds and/or other dynamic external fields, e.g., fluid separation[27], melt processing[28,29], nanomaterial synthesis[30,31], drug delivery[32,33] and targeted tumor treatment[34,35], energy conversion[36], pollution control[37], and many other applications.

## The experiment and operando measurements

We conducted a series of experiments at the 32-ID-B beamline of the Advanced Photon Source (APS) at Argonne National Laboratory (Fig. 1a). These experiments utilized two different storage ring filling modes for optimal image capturing and data acquisitions of different dynamic phenomena.

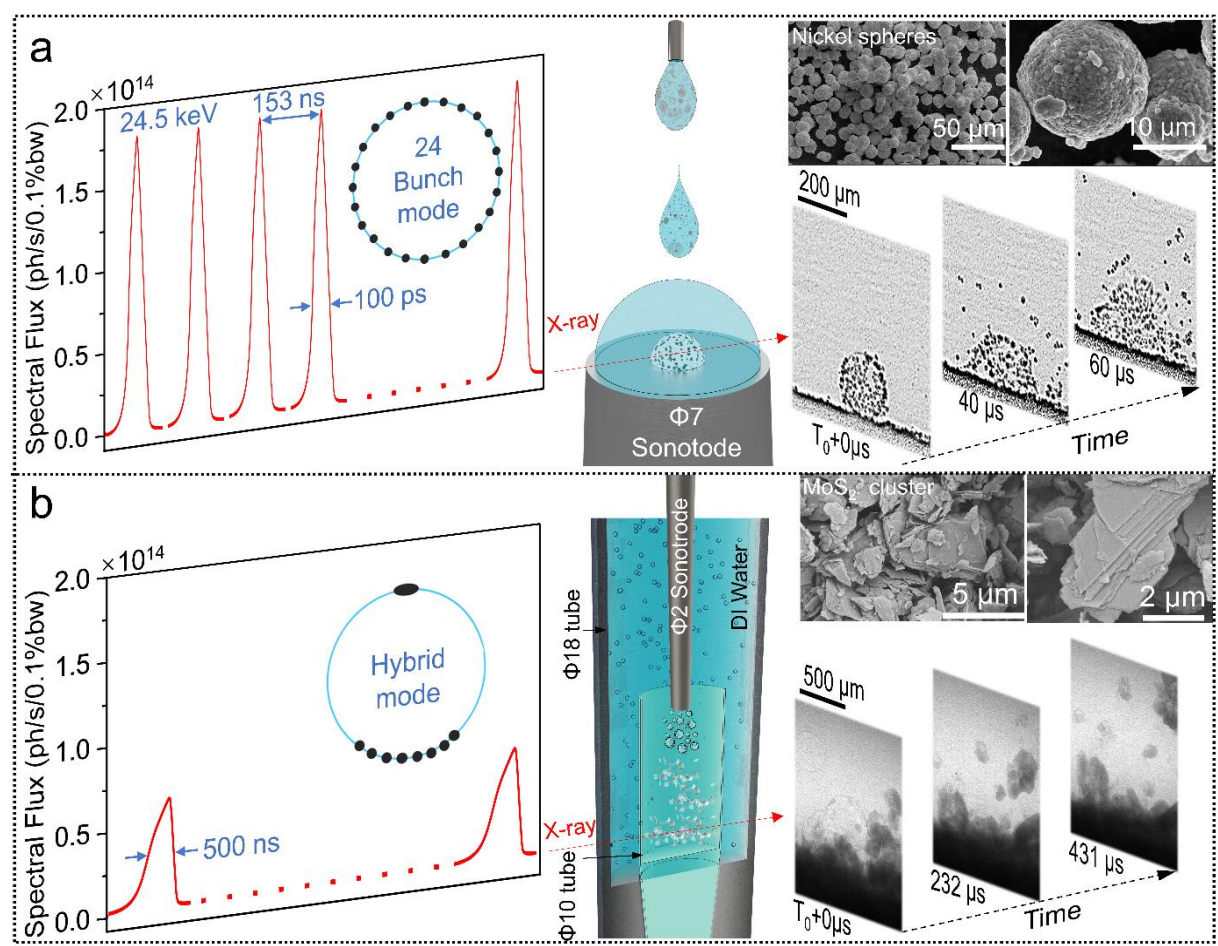

***Fig. 1. The experiment setup. (a)** From left to right: the X-ray pulses produced by the 24-bunch filling mode. The 7 mm diameter sonotrode tip (pointing upwards) and the Di-water drops that contained the Ni particles (delivered via a 1.3 mm diameter needle connected to a syringe pump). The scanning electron microscopy (SEM) images, showing the size and morphology of the spherical Ni particles. The typical X-ray images acquired at 100,000 fps, showing the movement of the Ni particles and the bubble interfaces. **(b)** From left to right: the X-ray pulses produced by the hybrid filling mode by the long 8 electron bunches. The 2 mm diameter sonotrode tip (pointing downwards) inside the 10 mm diameter tube that contained the $MoS_2$ particles (clusters). The whole assembly was immersed inside an 18 mm diameter tube filled with DI-water. The SEM images, showing the size and morphology of the $MoS_2$ particles. The typical X-ray images acquired at 30,173 fps, showing the dispersion dynamic of the $MoS_2$ clusters and the bubble implosion.*

Two specific experiment setups were used. Fig. 1a shows the use of a 7 mm diameter sonotrode (made by Ti-6Al-4V alloy), pointing upwards; and Fig. 1b shows the use of a 2 mm diameter sonotrode, pointing downwards. Both were driven by a Hielscher UP100H ultrasonic processor operated at a fixed frequency of 30 kHz. The amplitude was controlled by the input power. Using Fig. 1a setup without Ni particles, we firstly studied systematically the impact of fluid droplets on a vibrating surface, the atomization behaviour of a continuous fluid medium, and used the collected images to validate the model. Secondly, we mixed the Ni particles with deionized (DI) water and delivered the mixture onto the sonotrode surface via a syringe pump operated at a flow rate of 0.1 mL/min. Such experiment was deliberately designed for studying the interactions between Ni particles and ultrasonic bubbles.

Using Fig. 1b setup, we did systematic investigation on the ultrasound cavitation dynamics at different amplitudes and ultrasonic bubble interactions with micrometre particles, accounting for the effects of particle size, morphology, and surface properties (e.g., hydrophobic). Firstly, $MoS_2$ particles (clusters) were carefully placed into a 10 mm diameter quartz tube holder using a vacuum pick-up tweezer. Then, DI water was slowly flowed into the 18 mm diameter quartz tube container using a syringe pump without causing any disturbance of the $MoS_2$ particles before applying the ultrasound. Finally, the dynamic interactions between $MoS_2$ clusters and ultrasonic bubbles were imaged systematically.

**Bubble boundary evolution in ultrasound field and numerical simulation**

Fig. 2a presents a typical sequence of X-ray images, showing the evolution of a single bubble in water with the sonotrode pointed upwards. Initially, a bubble with a radius of ~27 μm was observed sitting on the sonotrode tip (Fig. 2a1). After the ultrasound was triggered, the spherical bubble surface began to distort at ~10 μs and evolves into a highly twisted surface morphology in ~60 μs (Figs. 2a2 to 2a5). Throughout this process, the bubble underwent stages of collapse (Fig. 2a6), C-shaping (Fig. 2a7), and microjet formation (Fig. 2a8). In all images, the bubble interfaces were clear and sharp because of the strong X-ray phase-contrast effect compared with images obtained by visible light illumination [38].

A comparison between the simulation (Fig. 2b) and the X-ray images (Fig. 2a) indicates that the simulation captured precisely the morphological evolution and the associated characteristics of the bubble shape at different stages, including the formation of a C shape at 165 μs and a microjet at 167 μs. The experimental results show that the bubble evolution was a few μs "lag behind" than those shown by the simulation results. This was because a constant and a steady-state acoustic pressure was used as the initial pressure condition in simulation (Fig. 2c). But in the experiment, there was a ramping up period for the sonotrode to reach the desired amplitude.

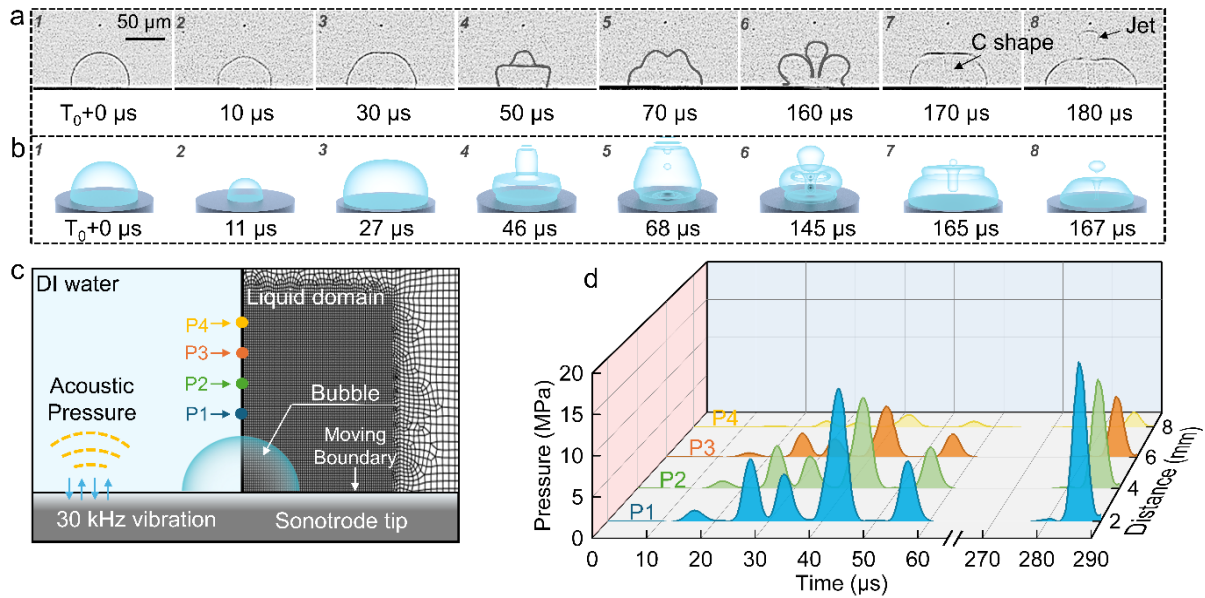

*Fig. 2. Bubble implosion dynamics in DI Water and model validation.* (a) A typical X-ray image sequence (acquired at 100,000 fps), capturing the implosion of a bubble on the sonotrode tip surface in DI water, with an input ultrasonic intensity of 44 W/cm². The series show the bubble distortion, collapse, C-shape formation, and microjet development. (b) The corresponding numerical simulation results, depicting the bubble dynamics over 8-10 oscillation cycles, showing excellent agreement with the experimental observations. Key features like the C-shape (at 165 μs) and microjet (at 167 μs) are accurately simulated. (c) The geometry, mesh structures and boundary conditions for the computational domain; (d) The calculated pressure profiles versus distance and time along the points P1~P4 marked in (c). (More dynamic information can be seen in Video 1).

During microjet formation, the maximum pressure and velocity can reach ~20 MPa (see P1 in Fig. 2d) and ~40 m/s (calculated from the liquid–gas interface movement shown by the X-ray images in Fig.2a), which are significantly higher than the ~1.6 MPa reported in previous studies where the bubble was located 1 mm away from the sonotrode (pointing upwards)[16]. This enhancement results from the direct contact between the bubble and the vibrating surface in the present study, which minimizes energy dissipation through the surrounding liquid.

**Bubble-particle interaction - the spherical particles**

To study the bubble-particle interaction dynamics, spherical hydrophobic Ni particles (see insert in Fig.1a) were used. Figs. 3a and 3b show the typical cases of bubble oscillation and bubble implosion, respectively.

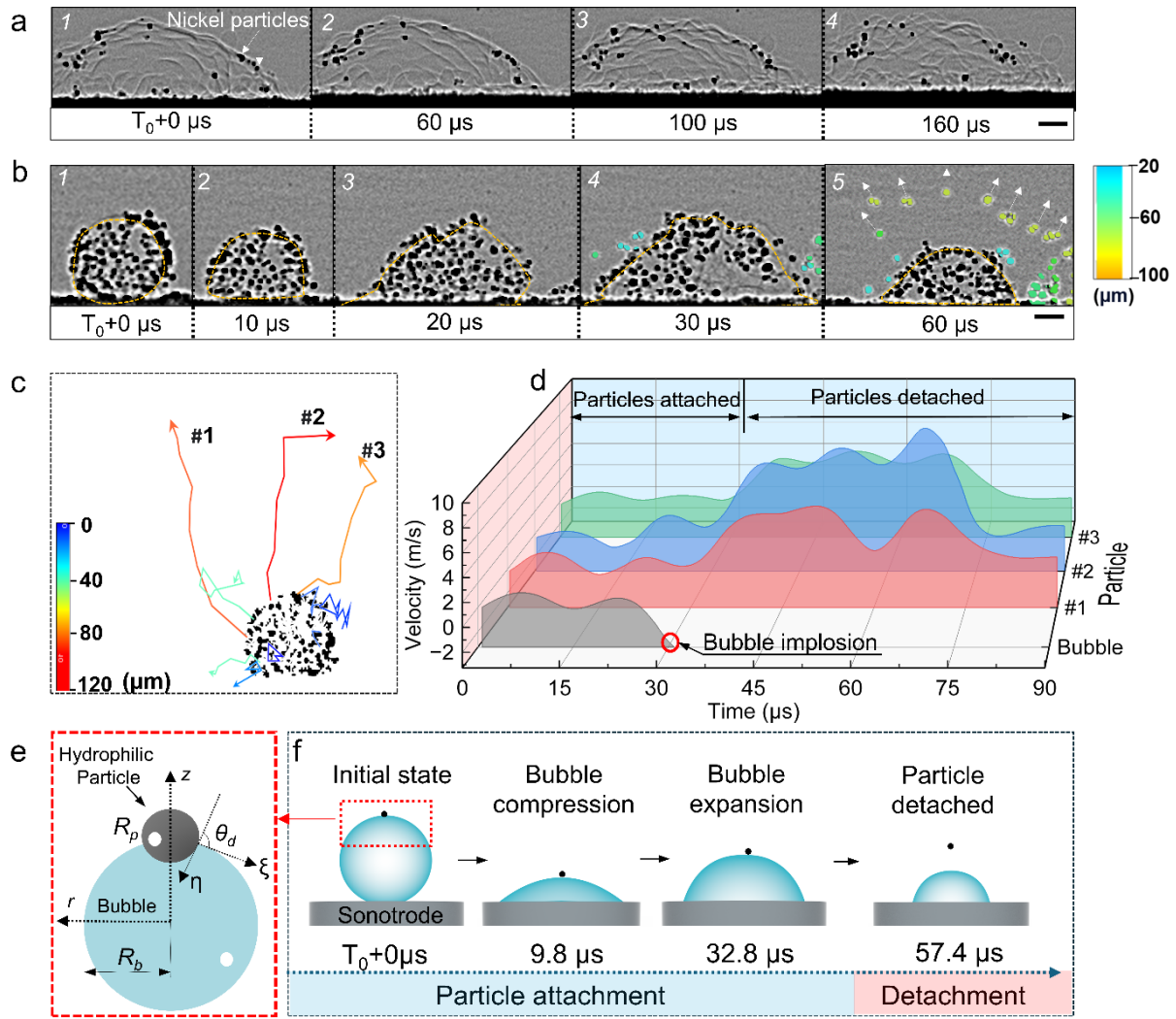

*Fig. 3. Dispersion dynamics of Ni spheres driven by oscillating and imploding bubbles.* (a) A typical sequence of X-ray images (50,000 fps), showing a few Ni particles attached to the bubble boundary which started to develop instability. (b) A typical sequence of X-ray images (100,000 fps), showing that the Ni particles initially covered on top of an oscillating bubble were ultimately dispersed at the bubble implosion. (c) The moving paths of three tracked Ni particles (colour-coded by displacement and tracked using the Track Mater plugin in Fiji) during the dispersion event. (d) Velocity profiles of the three particles and that of the bubble boundary (as the reference), showing when particles dispersion occurred. (e) and (f) The simulation results, showing the hydrophobic Ni particle interacted with an oscillating bubble. The scale bar is 50 μm. (More dynamic information can be found in Videos 2 and 3).

Fig. 3a illustrates the instability development of the bubble interface - a classical Rayleigh-Taylor instability[39] phenomenon. At this stage, despite the bubble interface started to deform and distort, the Ni particles remained attached. To quantify the dynamic movement of the particles, we applied a tractable modelling approach by simulating the interaction between a single Ni hydrophobic particle and a single bubble. Using this approach, we can calculate the energy transfer from a single bubble to a single particle, which allows us to precisely quantify the underlying physics and compare it with the theoretical calculations made by others, rather than to handle the highly complex scenarios of many particle systems. Fig. 3e shows the geometry setup (the equations and numerical schemes used are described in section 2 of the Supplementary Information). The particle's hydrophobic characteristics was described by the three-phase contact angle ($\theta_d$). In Fig. 3b, a bubble (its boundary is outlined by the yellow dotted line) covered with many Ni spherical particles was found at the sonotrode tip before the ultrasound was triggered. We used the TrackMate plugin[40] in Fiji (see Fig.3c) to track the moving trajectories of three typical particles from Fig. 3b, color-coded by their displacement. The results revealed that during bubble oscillation, particles also moved back-and-forth with the bubble interface and the average displacements were 8±3 μm. Clearly, kinetic energy transfer occurred when the particles were pushed outward at the bubble expansion period. At the same time, the surrounding liquid exhibited viscous resistance force to the movement of the bubble and particles, "helping to hold" the particles onto the bubble interface. After the particles gained sufficient energy (or velocity) through the above energy transfer stage (see Fig.3b3, 4), the particles were seen to continue to move outward due to the momentum gained (see Fig.3b4). When the bubble underwent a rapid contraction in the compression pressure period (see Fig.3b5), the particles were seen to "bounce off" the bubble boundary. Such detachment depends primarily on the adhesion force between the particles and the bubble boundary. The simulations shown in Fig. 3f captured precisely such dynamic behaviours, indicating that the simple and tractable modelling approach captured the underlying physics. The simulations showed that the oscillating speed of the particles induced by the bubble oscillation was 2~3 m/s (Fig.3d) and such speed was sufficient to make the Ni particles to bounce off (or detach) from the bubble boundary and hence to disperse into the surrounding liquid (Fig.3b5).

The X-ray images further revealed that, during bubble implosion, particles were immediately ejected due to the collapse of the bubble interface (the impact due to the local shock waves produced). The three tracked particles in Fig. 3c showed the ejected distance reached 47 ± 2 µm and the speed was ~10 m/s, i.e., 3–4 times higher than those in the oscillating bubble scenarios.

**Bubble-particle interaction - the flake particles**

Many two-dimensional (2D) layered materials, including graphite flakes and $MoS_2$ nanosheets, are hydrophobic and have lamellar morphologies, often promoting strong interlayer interactions and cluster formation (see the insert in Fig. 1b and Supplementary Information)[41]. Ultrasonic bubbles were often found to be "mingled" with the clusters, and it is very difficult for optical imaging to "see" clearly any dynamic interactions of such systems. In this aspect, ultrafast synchrotron X-ray imaging is an ideal technique to observe the dynamics of bubbles even within such clusters. Therefore, we conducted a series of experiments using $MoS_2$ clusters placed at different locations below the sonotrode tip. Three representative cases are presented below:

**(1) $MoS_2$ flakes dragged by quasi-statically oscillating bubbles**

Fig. 4a shows an X-ray image sequence of an $MoS_2$ cluster located ~12 mm below the sonotrode tip. From Fig. 4a2 to Fig. 4a3, the bubble expanded to a radius of ~35 µm during the rarefaction period. It then contracted to a radius of ~19 µm (Fig. 4a4) during the compression period.

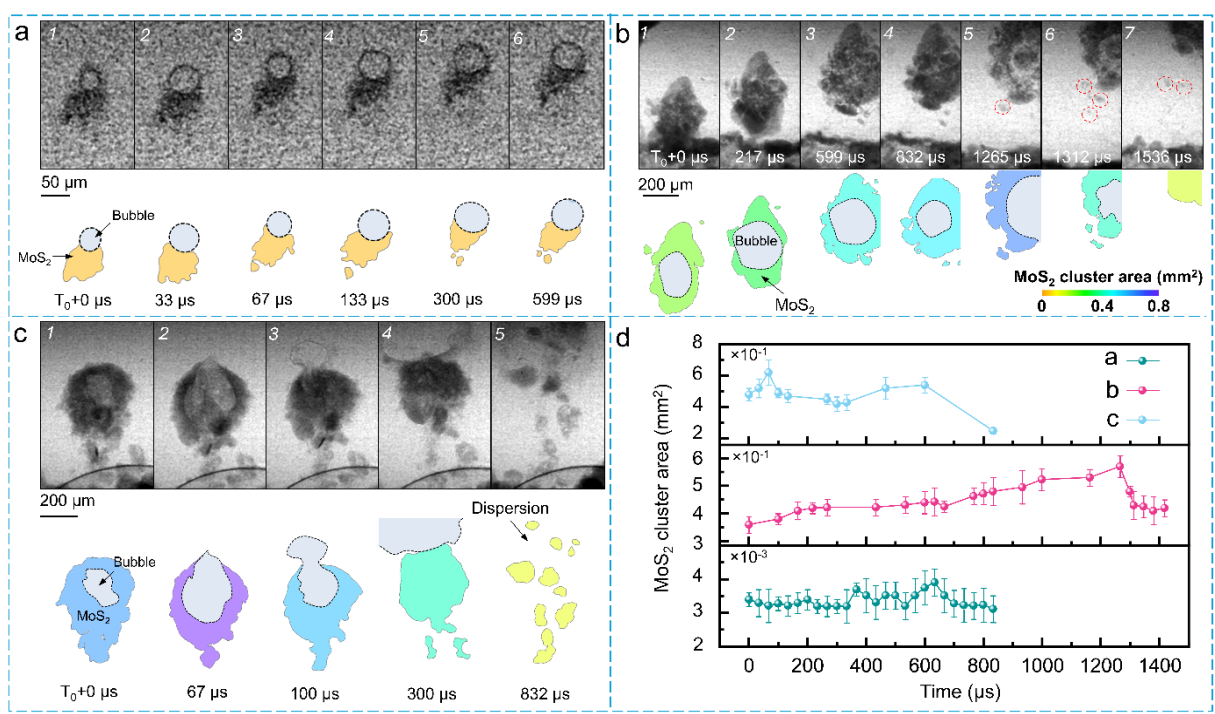

*Fig. 4. Interaction dynamics between ultrasonics bubbles and MoS$_2$ clusters. Three typical sequences of X-ray images captured at 30,173 fps at a location ~12 mm below the sonotrode tip, showing (a) a small cluster of hydrophobic MoS$_2$ particles partially attached to an oscillating bubble. (b) A bubble nucleated within a loose MoS$_2$ cluster, became fully covered by the cluster. The cluster was gradually dispersed through bubble oscillation and expansion. (c) A bubble expanded and escaped from a loosely MoS$_2$ cluster, collapsing and instantly fragmenting the cluster into smaller pieces. In all cases, the bubbles and surrounding clusters are segmented and quantified using a machine learning algorithm, with the corresponding quantitative segmented images displayed below each set of X-ray images. (d) Evolution of the cluster area over time for the above three cases shown in (a) to (c), quantifying the dispersion process. (More dynamic information can be found in Videos 4-6).*

As the bubble oscillated and moved upward, a small cluster of MoS$_2$ particles attached at its lower boundary and was dragged up together with the bubble. Image analysis indicated that the bubble moved at a velocity of ~ 0.3 mm/s, while the estimated mass of the MoS$_2$ cluster moving with the bubble was ~3.4 µg. This phenomenon is not intuitive given that the bulk density of MoS$_2$ (5.06 g/cm³) would typically cause it to sink in water to the bottom of the tube. This observation clearly demonstrated that (1) the "attractive" force exerted by the bubble boundary on the MoS$_2$ cluster and (2) the buoyancy provided by the oscillating and upward-moving bubbles collectively overcame the gravitational force acting on the cluster. Furthermore, the MoS$_2$ particles were found to cover and wrap the bubble surface completely.

## (2) Nucleation of bubbles in the voids between MoS$_2$ flakes

Fig. 4b illustrates an image sequence captured at a similar location (~12 mm below the sonotrode tip). In Fig. 4b1, a bubble was observed to nucleate and grow within a loose cluster of MoS$_2$ particles (a typical case where optical imaging would not give clear images). As the bubble oscillated and moved upward, its entire surface was covered with MoS$_2$ particles. Video 3 clearly shows the movement of this "bubble-particles" mixture with the oscillating bubble remaining fully covered by the MoS$_2$ particles. As the bubble oscillation continued, the whole area of the MoS$_2$ wrapping around the bubble was observed to expand from 0.18 mm² in Fig. 4b2 to 0.23 mm² in Fig. 4b5. This expansion indicates a progressive increase in inter-particle spacing, demonstrating that the clusters were effectively dispersed by the bubble oscillation and expansion. Interestingly, in Fig. 4b5-b7, some fragments (highlighted in red circle) had already detached from the cluster, while others remain stuck to the bubble interface until the bubble moved out of the FoV at the upper-right corner.

**(3) Bubble implosion and dispersion of MoS$_2$ flake clusters**

Fig. 4c shows a more violent interaction between a bubble and a MoS$_2$ cluster. In this case, the bubble interfaces vibrated more chaotically and lost its spherical shape, becoming high distorted with regions of sharper local curvature (both concave and convex shape with respect to the surrounding cluster). Video 4 clearly shows that such energetically and violently vibrating bubble is much more effective in dispersing the agglomerated MoS$_2$ clusters. More interestingly, the inward-distorted (concave) regions of the bubble boundary trapped some individual particles, which were subsequently dragged and entrained inside the bubble, as shown in Figs. 4c2 to c5 (corresponding to 865.8 µs to 965.7 µs in Video 4). Sufficient boundary distortion eventually caused the original bubble-particle cluster to disintegrate completely, producing smaller bubble debris which jumped out of the original cluster, thereby dispersing the particles more effectively. The following section gives a quantitative analysis of how energy transfer from bubbles to the particles occurred at bubble oscillation and implosion.

**Energy transfer during bubble-particle interactions and dispersion dynamics**

Previously, some attempts for calculating energy transfer in multiphase flows were made using analytical models with oversimplified assumptions[42] because a complete set of numerical models that can calculate the multiple spatial and temporal scale phenomena is, in most cases, computationally prohibitive[43]. At microscale, events such as bubble collapse and violent interfacial instabilities occur in sub µs or shorter, requiring ns temporal resolution, i.e., time step of ns or shorter[18,28,44]. In contrast, macroscopic flow evolution and acoustically driven energy redistribution occur in seconds to minutes[18]. Applying ns time step globally in the computational domain needs billions of time steps in a few seconds simulation, making such simulations prohibitively expensive and sometimes highly susceptible to the accumulated numerical errors.

To overcome such multi-scale challenge, we developed a hybrid analytical–numerical framework. In this approach, the transient bubble dynamics are resolved using high-fidelity compressible flow simulations (see section 2 of the Supplementary Information), validated against the synchrotron X-ray images acquired, produced time-resolved information (profiles) for the bubble volume, internal pressure, density, and temperature. The resulting microscale data were then incorporated as dynamic source terms into an analytical model—derived from volume-averaged conservation laws, which is highly computationally efficient in calculating the macroscopic, quasi-steady state energy transfer. The resulting tightly coupled scheme reduces computational overhead by ~5 orders of magnitude compared to fully resolved numerical simulation schemes, while preserving microscale accuracy, enabling robust and quantitative prediction of energy partitioning across spatiotemporal scales.

We used Eq. 1-3 below to calculate the energy balance in a mesoscale bubble-particle system. The total local energy ($E_L, Eq.1$) [45] consists of local kinetic energy ($E_K, Eq.2$) and local potential energy ($E_P, Eq.3$). As a bubble expands or collapses adiabatically from its initial volume ($V_0$) to its current volume ($V$), the local potential energy ($E_P$) equals the sum of (i) the work done on the liquid by the gas pressure at the bubble wall and (ii) the energy transmitted across the boundary to surrounding particles. The local kinetic energy ($E_K$) is contributed by the liquid flow induced by bubble motion, since the gas density is typically three orders of magnitude lower than that of the liquids.

$$E_L = E_P + E_k \tag{1}$$

$$E_k = \frac{p_g V_0}{\kappa - 1}\left(\frac{V_0}{V}\right)^{k-1} + \sigma A - \rho g V Z_c \tag{2}$$

$$E_P = (p_\infty - p_v)V + \frac{1}{2}p_\infty \oint_S \varphi \varphi_n \, dS \tag{3}$$

$$E_T = E_{L1} - E_{L2} \tag{4}$$

where $V_0$ and $V$ are the initial and final bubble volumes, $g$ is the acceleration of gravity, $\sigma$ is the surface tension coefficient, the reference pressure is defined as $\Delta p = p_\infty - p_v$, where $p_\infty$ is the pressure in the undisturbed liquid and $p_v$ is the vapor pressure inside the bubble ($p_v \sim 2$ kPa at 20°C)[16], $A$ is the areas of the current bubble surfaces, $\kappa$ is the polytropic index of the bubble gas, $\varphi$ and $\varphi_n$ is the velocity potential and its normal derivative on the bubble surface S, respectively. $E_{L1}$ and $E_{L2}$ in Eq.4 represent the local energy of bubbles with and without a particle, respectively, and their difference $E_T$ is the local energy transferred from the bubble to the attached particle.

The fluid dynamics and bubble dynamics numerical models, which were validated against the extensive X-ray datasets, provide the key information such as pressure ($p$), the velocity potential $\varphi$ and normal velocity $\varphi_n$ on the bubble interface, which are necessary to calculate $E_L$. By coupling the analytical and numerical models, we can calculate the kinetic energy transfer, local potential energy changes and energy loss to surrounding particles during bubble oscillation and implosion, through a direct comparison of the local energy without ($E_{L1}$) and with ($E_{L2}$) particles.

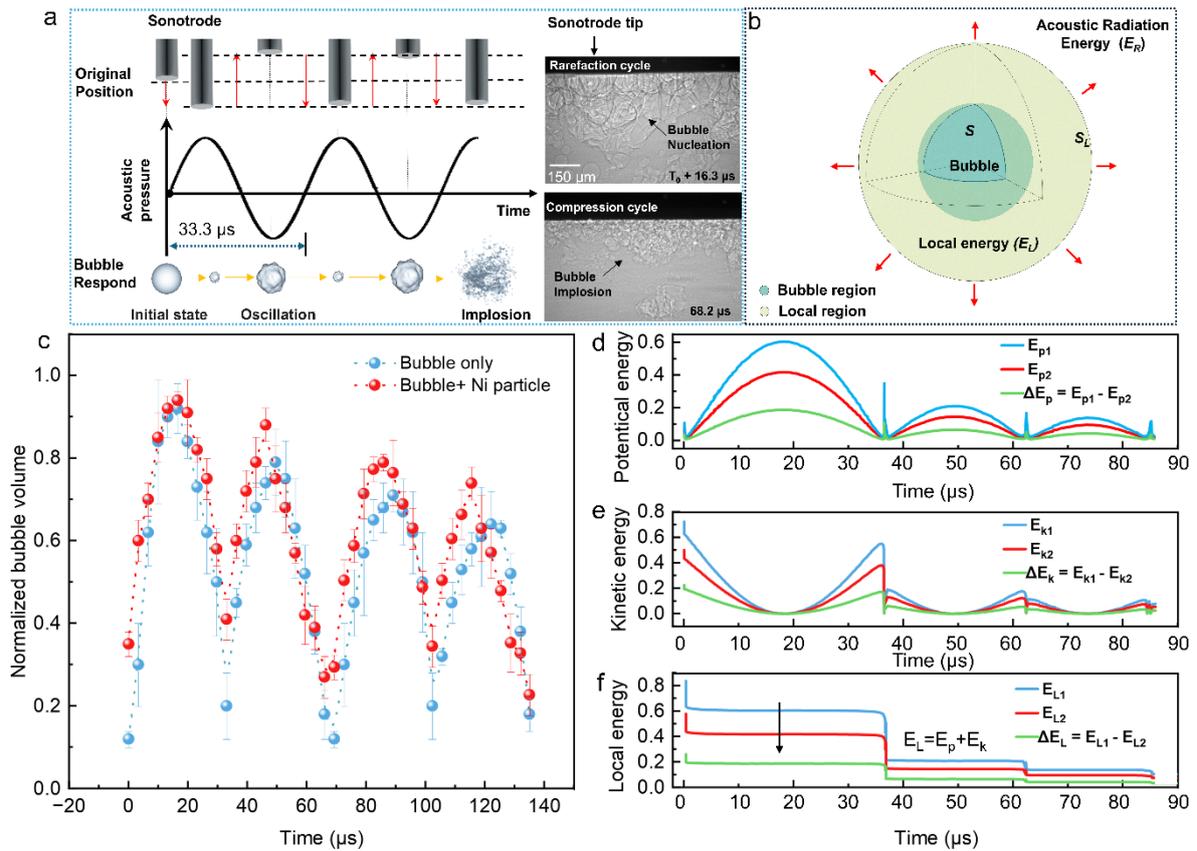

*Fig. 5. Energy transfer induced by bubble oscillation.* (a) The sonotrode positions with time, accompanied by the corresponding acoustic pressure wave and the bubble response. (b) Illustration of the local region (Ω) of the flow field surrounding a bubble in an infinite domain, bounded by the bubble surface ($S$) and a large sphere ($S_L$). The total energy of the bubble system consists of the local energy ($E_L$) within the inner region and the acoustic radiation energy ($E_R$) emitted outside the inner region. (c) The normalized bubble volume over two ultrasound cycles for the two cases. The blue lines (dotted and solid) represent the bubble-only system, while the red lines (dotted and solid) represent the bubble + Ni particles system. The dots with error bars indicate the experimental measured bubble radius from the X-ray images. The corresponding (d) potential energy and (e) kinetic energy profiles of the bubble system without (blue line) and with particles (red Line) are normalized by the initial potential energy $E_{p0}$ and the initial local energy $E_{k0}$, the difference between these profiles (green line) reflects the energy transferred to the liquid and particles driven by bubble oscillation. (f) Local energy (sum of potential and kinetic energy) profiles and associated energy transfer for the system with and without particles.

Fig.5c shows the bubble volume (normalised) evolution during oscillation over two ultrasound periods (see Fig. 5c, and the corresponding X-ray images are from Fig.S4 in Supplementary information). Based on these data, we calculated the corresponding potential (Fig. 5d) and kinetic energy profiles (Fig. 5e) to illustrate how the oscillating bubble transfers energy to the liquid and particles. As the bubble volume began to increase, energy was predominantly transferred to the particle as kinetic energy, with only a negligible fraction dissipated into the liquid. As the bubble continued to expand, a portion of the energy was converted and transferred to the liquid and particle as potential energy. Once the bubble reached its maximum volume (e.g., 18 µs in Fig. 5d,e), all kinetic energy was converted into potential energy and delivered to the liquid and particles. This process reversed during the contraction phase.

Fig.5f presents the temporal evolution of the local energy of the bubble volume over three consecutive cycles. The bubble underwent several oscillation cycles, with each cycle's amplitude decreased due to local energy loss. The local energy remained almost constant in each specific oscillation cycle but decreased abruptly at the minimum volume value at the end of that cycle. When a hydrophobic particle was present on the bubble surface, the overall profile of the local energy was similar to the case without particle, but local energy loss increased by up to 16% (compare the red and blue solid lines in the first cycle). Such additional dissipation was transferred to the surrounding liquid and particles through bubble oscillation, promoting the dispersion of agglomerated particles around the bubble.

Under the present experimental conditions (initial bubble radius in the range of $R_0$ = 50~150 µm, driving frequency f = 30 kHz, acoustic pressure amplitude 1.5 MPa), the total energy of a single oscillation bubble is estimated to be 0.5–2 µJ using the Eq.5 below derived from the Rayleigh–Plesset equation:

$$E_{osc} \approx 4\pi\rho R_0^3 \left(\frac{P_a}{\rho_w R_0}\right)^2 \quad (5)$$

where, $E_{osc}$ is the energy of bubble oscillation, representing the characteristic energy associated with the oscillatory motion of the bubble or interface. $R_0$ is the characteristic (or equilibrium) radius of the bubble, $P_a$ is the amplitude of the applied acoustic pressure, $\rho_w$ is the density of DI water.

Thus, the energy delivered to a single particle is

$$E_T = \eta \times E_{osc} \tag{6}$$

where $\eta$ is the energy transfer efficiency from the oscillating bubble to the particle, which is approximately 16% estimated in the first cycle in Fig.5f. Therefore, the energy transferred to a single particle is 80–320 nJ. Assuming the single particle is $MoS_2$ (with lateral dimensions 1~7 μm), the energy required for dispersion or partial exfoliation is primarily governed by the interlayer van der Waals binding energy. High-precision *in situ* peeling-to-fracture measurements have determined this interlayer binding energy to be 0.55 ± 0.13 J m⁻² [46]. Scaling this value to realistic flake areas and incorporating corrections for multi-layer aggregation and ultrasonic energy transfer efficiency yields an estimated dispersion threshold per particle of 20–80 nJ. The estimated energy transfer indued bubble oscillation in our case exceeds the dispersion threshold, which is fully consistent with our experimental observations: as clearly demonstrated in Fig. 4b, efficient detachment and dispersion of $MoS_2$ cluster are readily achieved under the applied acoustic conditions, indicating that the delivered energy exceeds the required threshold for effective particle exfoliation.

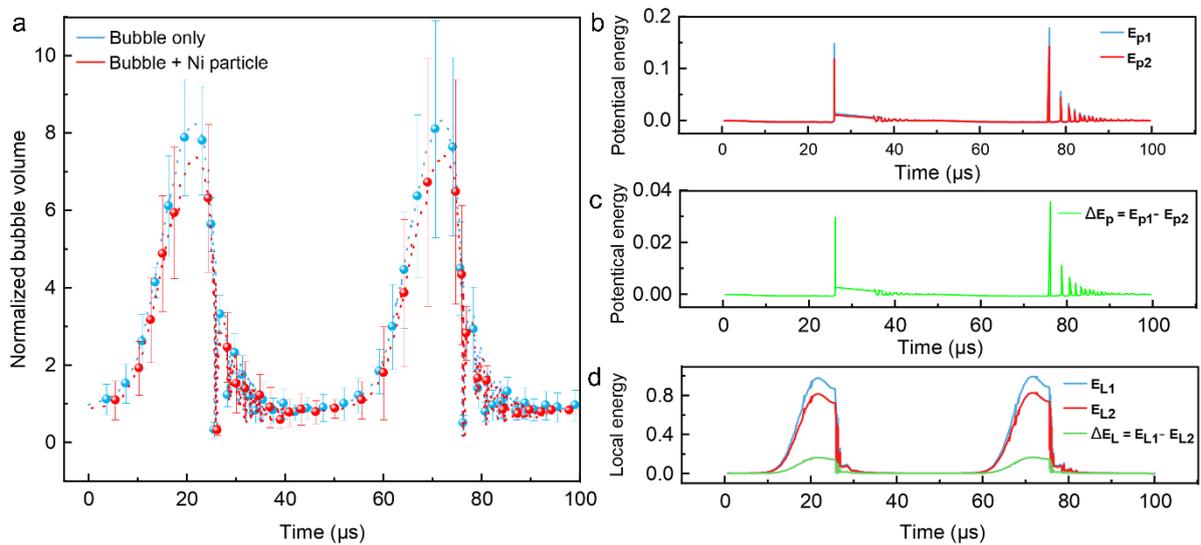

***Fig.6. Energy transfer induced by bubble implosion.*** *(a) The normalized (experimentally measured) bubble volume over two ultrasound cycles for bubble implosion. The blue legends (dotted line and big solid dots) represent the bubble only. The red lines are for the bubble + Ni particles. The big dots with error bars show the measured bubble volume from the x-ray images. The corresponding (b) & (c) potential energy and (d) local energy profiles of the bubble system with and without particles are normalized by the initial potential energy $E_{p0}$ and local energy $E_{k0}$. The difference between these profiles (green line) reflects the energy transferred to the liquid and particles driven by bubble implosion.*

Based on the normalized bubble volume shown in Fig. 6a (the corresponding X-ray images are from Fig.S4 in Supplementary information), we used the same approach to calculate the local and potential energy transfer between a bubble and a particle during bubble implosion. As shown in Fig. 6d, the local energy increased rapidly from near zero to its peak in a very short period (~14 μs, or 30% of the acoustic cycle). It then decayed even more rapidly to near zero in ~5 μs (8% of the cycle) and remained low for the rest of the cycle. Such ultrafast energy exchange drove the bubble into violent oscillations, expanding to over eight times its equilibrium radius before imploded in ~4 μs. After the imploded, the bubble continued to oscillate with small amplitude. Our calculations showed that the energy transfer induced by bubble implosion accounted for approximately 26% of the total energy transferred to the surrounding particle. Such energy transfer can be decomposed into two contributions: (i) the particle's initial kinetic energy gained from bubble oscillations before implosion, and (ii) the particle's potential energy transferred by the shockwave during bubble implosion. Each transfer occurred almost simultaneously and completed in sub- μs (Fig. 6b,c). The energy released during bubble implosion (inertial cavitation) can be described as the conversion of the bubble's potential energy into kinetic energy during its violent collapse. Following the classical Rayleigh model, this energy can be approximated as:

$$E_{implosion} \approx \frac{4}{3}\pi\rho R_{max}^3 P_\infty \qquad (7)$$

Using the similar approach as for the case of bubble oscillation, the energy transferred to the particles when bubble collapse (inertial cavitation) was in a range of 0.135–1.09 μJ (assuming 26% transfer efficiency via shock waves, microjets, and shear forces during violent implosion). Although the current model considers only a single hydrophobic particle, the energy transfer between a bubble and multiple particles can be quantified by considering the particles-to-bubble surface area ratio, i.e., the coverage[47], and therefore extended to a "single bubble + multiple particles" system. This scalable approximation is technically significant for quantitatively describing the multi-particle scenarios observed in Figure 3b.

In summary, we used ultrafast synchrotron X-ray imaging to capture, in operando conditions at μs time and μm length scale, the dynamic interactions among ultrasonic bubbles and hydrophobic particles (clusters). In particular, the dynamic interactions in the whole process of cavitation bubble nucleation, oscillation, and implosion. Key dynamic information was extracted from the X-ray images and then directly fed into (or informed) the hybrid analytical-numerical framework we developed. Using both approaches synergistically, we are able to quantify accurately the energy transfer from ultrasonic bubble to the particles at bubble oscillation and implosion. Our research indicated that for the hydrophobic Ni spherical microparticles, at bubble oscillation, ~16% (80–320 nJ) of the bubble interface energy is transferred the particle, while at implosion, ~26% (0.135–1.09 µJ) to the particle. The transferred energy is mainly consumed by particle dispersion. Our research provides valuable insights and mechanisms for guiding the ultrasound dispersion or processing of hydrophobic micro and nanomaterials, which is applicable to a wide range of scientific and industrial processes involving interfacial energy transfer, such as making suspensions, composite materials and exfoliated 2D materials.

**Methods**

**Ultrasound processing setup and experimental procedure**

For the ultrasound experiments, a Hielscher UP100H ultrasonic processor was used to drive the sonotrodes (all made of Ti-6Al-4V alloy) of either 7 mm diameter (Fig.1a) or 2 mm diameter (Fig.1b). The ultrasound frequency was 30 kHz and the amplitudes were determined by the input powers. In Fig. 1a, the sonotrode tip was pointing upwards. Liquid drops of deionised water containing Ni microparticles (99.8% purity, 10~15 µm in diameter, purchased from Alfa) were delivered directly onto the sonotrode tip surface via a 1.3 mm diameter needle connected to a syringe pump (Darwin NE-1000). The flow rate used was 0.1 ml/min. The Ni microparticles (10 grams) were premixed uniformly with deionised water (50 ml) inside a glass beaker for 5 minutes by magnetic stirring. In the experiments using the $MoS_2$ clusters, the sonotrode tip was pointing downwards (Fig. 1b). The $MoS_2$ clusters (~11g, 99% purity, 6~10 µm flakes, purchased from Nanoshel) were carefully filled into a small quartz tube holder (10 mm O.D., 0.5 mm wall thickness) using a vacuum pick-up tweezer. Then ~11 ml DI water was gently "flowed" into a larger quartz container (18 mm O.D., 0.5 mm wall thickness, 80 mm tall) along its inner wall using the syringe pump.

## In situ synchrotron X-ray radiography acquisition

The experiments were conducted at the 32-ID-B beamline of the Advanced Photon Source (APS) at Argonne National Laboratory, USA, using two distinct storage ring filling modes. For the experiment depicted in Fig. 1a, the 24-electron bunch filling mode was used. The undulator gap was set to ~12 mm, generating a white X-ray beam with its first harmonic of ~ 24.5 keV, a pulse width of ~100 ps, and a peak spectral flux of ~1.8×10$^{14}$ photons per second per 0.1% bandwidth. The time interval between consecutive X-ray pulses was ~153 ns. Theoretically, an image acquisition rate of up to 6.5 MHz frames per second (fps) is possible if every individual X-ray pulse is used for imaging. We selected acquisition rates of 100 kHz and 50 kHz to capture images over a longer duration, aligning with the time scale of the physical phenomena under investigation. In the experiment shown in Fig.1b, the hybrid electron filling mode (a long train of 357.64 ns with 8 electron bunches plus a short single electron bunch of 1.6 ns) was used instead. The undulator gap was set at ~12.5 mm, and the long 8 electron bunch train produces a white beam with the first harmonic (~500 ns width) at ~24.4 keV and the peak spectral flux of ~6.17 × 10$^{13}$ ph/s/0.1%bw. In this mode, the highest possible image acquisition rate is ~271,554 fps. However, in practice, the FoV of the Photron camera is inversely related to the acquisition rate. For this experiment, we selected an acquisition rate of 30,173 fps with an exposure time of 1 μs, which provided a sufficiently large FoV to observe the $MoS_2$ clusters effectively.

For X-ray imaging in both setups, the LuAG: Ce scintillator (70 ns delay time) coupled with a 10× objective lens was used. The sample-to-detector distance was set at ~290 mm. A Shimadzu HPV-X2 camera was used in Fig. 1a and the full field of view (FoV) was 1250×750 μm$^2$ (3 μm/pixel). A Photron FastCam SA-Z camera was used in Fig.1b with variable FoV dependent on the acquisition rates. The effective spatial resolution was 1.9 μm/pixel.

## Microstructure characterization

SEM images were acquired using a FEI Helios Nanolab G3 in secondary electron (SE) mode with an Everhart–Thornley detector. All samples were carbon-coated prior to imaging, and measurements were performed at 25 kV with a beam current of 50 pA to enhance contrast and minimize charging.

**Hybrid analytical–numerical modelling**

A hybrid analytical–numerical modelling approach was used to calculate multi-time-scale energy transfer with the key parameters extracted directly from the ultrafast X-ray images obtained. The numerical simulation of ultrasonic bubble dynamics was carried out using the commercial software ANSYS Fluent 2021 R2. The fully coupled governing equations, including continuity, momentum, and energy conservation, were solved using the Finite Volume Method (FVM). To model bubble dynamics under ultrasound, the Volume of Fluid (VOF) and Continuous Surface Force (CSF) methods were used to simulate bubble oscillation and implosion under ultrasound. The model was calibrated and validated using X-ray image data obtained from imaging experiment. For the interaction between particles and bubble, the motion of solid particles was simplified by considering only a single solid particle moving in the vertical direction and interacting with oscillating/imploding bubbles. For further details and parameter specifications, please refer to the Supplementary information.

**Acknowledgments**

The authors would like to acknowledge financial support from the UK Engineering and Physical Sciences Research Council (Grant Nos. EP/R031665/1; EP/R031401/1; EP/R031819/1; EP/R031975/1). This research used resources of the Advanced Photon Source (Proposal ID: 72860); a U.S. Department of Energy (DOE) Office of Science User Facility operated for the DOE Office of Science by Argonne National Laboratory under Contract No.DE-AC02-06CH113